\documentclass[reprint,nofootinbib,amsmath,amssymb,fontenc aps]{revtex4-1}
\usepackage{graphicx,dcolumn,bm,hyperref,float}
\usepackage[shortlabels]{enumitem}
\usepackage{color}
\usepackage{anyfontsize}

\newcommand{\be}{\begin{equation}}
\newcommand{\ee}{\end{equation}}
\newcommand{\beq}{\begin{equation}}
\newcommand{\eeq}{\end{equation}}
\newcommand{\ba}{\begin{eqnarray}}
\newcommand{\ea}{\end{eqnarray}}
\newcommand{\bef}{\begin{figure}}
\newcommand{\eef}{\end{figure}}
\newcommand{\amp}{&\!\!\!}

\newcommand{\p}{\partial}

\newcommand{\cO}{{\cal O}}

\newcommand{\cH}{{\cal H}}

\def\q{{\bf q}}
\def\x{{\bf x}}

\def\k{{\bf k}}

\def\pv{{\bf p}}

\def\bv{{\bf b}}
\def\Lv{{\bf L}}

\newcommand{\ket}[1]{\left|#1\right\rangle}
\newcommand{\braket}[2]{\left\langle {#1} \middle| {#2} \right\rangle}

\newcommand{\rg}{\sqrt{-g}}

\newcommand{\A}{V^{(0)}}
\newcommand{\B}{V^{(2)}}

\newcommand{\sig}{\bar{\sigma}}

\newcommand{\scal}{\chi}

\newcommand{\Eq}{E_q}
\newcommand{\psik}{{\tilde{\psi}_\k}}
\newcommand{\F}{F}
\newcommand{\DM}{\text{DM}}
\newcommand{\Gamd}{\Gamma_{\mbox{\tiny{dec}}}}

\newcommand{\sigmat}{\tilde{\sigma}}

\newcommand{\qtv}{{\bf\tilde{q}}}

\newcommand{\oma}{\omega}
\newcommand{\scprof}{\phi_s}
\newcommand{\ma}{m_a}
\newcommand{\rhoDM}{\rho_{\mbox{\tiny{DM}}}}

\begin{document}

\title{General Relativistic Decoherence with Applications to Dark Matter Detection}

\author{Itamar J.~Allali$^*$}
\author{Mark P.~Hertzberg$^\dagger$}
\affiliation{Institute of Cosmology, Dept.~of Physics and Astronomy, Tufts University, Medford, MA 02155, USA}

\begin{abstract}
Quantum mechanics allows for states in macroscopic superpositions, but they ordinarily undergo rapid decoherence due to interactions with their environment. A system that only interacts gravitationally, such as an arrangement of dark matter (DM), may exhibit slow decoherence. In this Letter, we compute the decoherence rate of a quantum object within general relativity, focusing on superposed metric oscillations; a rare quantum general relativistic result. For axion DM in a superposition of the field's phase, we find that DM in the Milky Way is robust against decoherence, while a spatial superposition is not. This novel phase behavior may impact direct detection experiments.
\end{abstract}

\maketitle

{\em Introduction}.---
Matter that interacts infrequently can maintain its quantum coherence, making it an excellent candidate for rich quantum mechanical behavior. Such behavior includes macroscopic superpositions of distinctly observable states, which can give rise to exotic phenomena such as interference.
The potential for this behavior is encoded in the off-diagonal elements of the density matrix describing the object's quantum state. Typically, the quantumness of ordinary matter is destroyed rapidly due to \textit{decoherence}, a process by which interactions with the environment effectively suppress the off-diagonal elements of the density matrix \cite{Zeh1970,Zurek1981,Zurek1982,Joos:1984uk,Gallis1990,Diosi1995,Giulini:1996nw,Kiefer:1997hv,Dodd:2003zk,Hornberger,Schlosshauer:2003zy,SchlosshauerBook}.
Primarily gravitationally interacting matter, which we refer to as dark matter (DM),
exists somewhat isolated from its environment. Thus, if a piece of DM were to form a macroscopic superposition,
it may preserve its quantum coherence for 
macroscopically
long periods of time. This can have ramifications for direct detection as we develop in this work for the first time.

In this Letter,
we study a localized mass distribution of \textit{Dark Matter in a Superposition of Macroscopic States (DMSMS)}
and compute the rate of decoherence induced by general relativistic scattering of surrounding standard model (SM) particles. The formalism and results we obtain rely on genuinely relativistic effects in the weak-field metric approximation. For an analysis based on Newtonian gravity see \cite{Allali:2020ttz} and for a detailed relativistic companion paper see \cite{Allali:2020shm}. 
What we develop here is a truly {\em quantum general relativistic result}, all treated rigorously within effective field theory, as the effects arise from the metric $g_{\mu\nu}$ in a quantum superposition. 
There exist very few robust quantum general relativistic results; perhaps the only known examples are Hawking radiation \cite{Hawking:1974sw} and corrections to the gravitational force law \cite{Donoghue:1993eb,Ford:2015wls}.
Other work where decoherence and gravity or cosmology play some role, includes \cite{Bassi:2017szd,Belenchia:2018szb,Asprea:2019dok,Anastopoulos:2013zya,Blencowe:2012mp,Breuer:2008rh,Shariati:2016mty,DeLisle:2019dyw,Orlando:2016pwg,Pang:2016foq,Oniga:2015lro,Bonder:2015hja,Diosi:2015vra,Colin:2014vfa,Hu:2014kia,Pikovski:2013qwa,Polarski:1995jg,Halliwell:1989vw,Kiefer:1998qe,Padmanabhan:1989rm,Kafri:2014zsa,Nelson:2016kjm,Anastopoulos:2014yja,Wang:2006vh,Kok:2003mc,Pikovski:2015wwa,Kiefer:1999gt,Mavromatos:2007hv,Tegmark:2011pi,Anastopoulos:1995ya,Colin:2014cfa,Kiefer:2008ku,Brandenberger:1990bx,Khosla:2016tss,Podolskiy:2015wna,Arrasmith:2017ogi,Albrecht:2018prr}.
As we explain, our work is relevant to direct detection searches for the axion.

\begin{figure}[t]
\centering
\includegraphics[width=1\columnwidth,height=2.3cm]{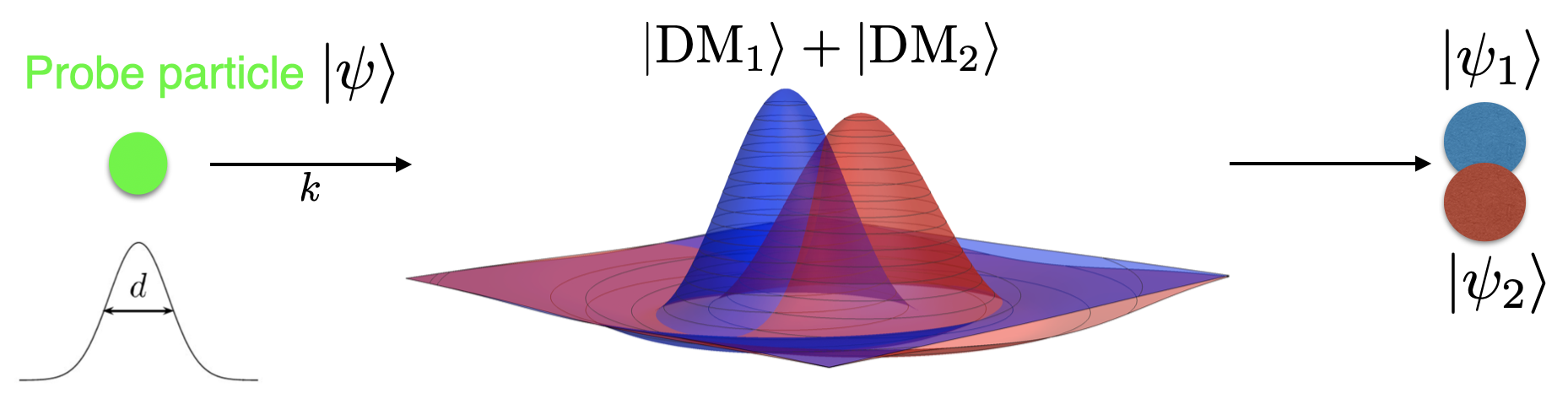}
\caption{A schematic representation of a probe particle scattered by Dark Matter in a Superposition of Macroscopic States (DMSMS),
evolving into an entangled state $|\Psi\rangle$ (see eq.~(\ref{ent})) where the probe particle's state is altered by gravitational interaction.}
\label{scheme} 
\end{figure}

{\em Basic set up}.---
Let us consider a 
DMSMS
which is some distribution of DM, represented pictorially in fig.~\ref{scheme} and described by a state $|\mbox{DM}_1\rangle+|\mbox{DM}_2\rangle$. The environment consists of probe particles, described by a quantum state $\ket{\psi}$, and the entire state $|\Psi\rangle$ evolves under the Schr\"odinger equation. 
Upon scattering the particle becomes entangled with the DM due to the gravitational interaction into the state
\beq
|\Psi\rangle=|\psi_1\rangle|\mbox{DM}_1\rangle+|\psi_2\rangle|\mbox{DM}_2\rangle
\label{ent}\eeq
By tracing out the probe particle, one can study decoherence. 
For simplicity, we are considering a superposition of only two states, $\ket{\DM_1}$ and $\ket{\DM_2}$. This basic setup can be easily extended to a more general superposition, which we comment on later. 
To proceed, we need the Hamiltonian $\hat{H}$ governing the gravitational interaction provided by general relativity. Throughout this work, we set $\hbar=c=1$.

{\em Hamiltonian and scattering}.---
To construct the desired Hamiltonian, we treat the probe particle as a scalar, ignoring its spin. Of course, realistic probes, such as baryons and photons, are not scalars, but since we are only interested in the leading universal gravitational interaction,
we leave the analysis of spin effects for possible future work. The gravitational interaction is incorporated in a local way by using the field formalism (with the probe taken to be a complex scalar field for convenience, though we will be interested only in particle states and not anti-particle states). 

The action for a complex scalar probe field $\scal$ minimally coupled to the metric $g_{\mu\nu}$ (signature $+---$) is given by
\beq S=\int d^4 x \sqrt{-g}\left(g^{\mu\nu}\p_\mu \scal^*\p_\nu\scal - m_p^2 \scal^*\scal\right)
\eeq
where $g$ is the metric determinant, $g^{\mu\nu}$ is the inverse metric, $\scal^*$ is the complex conjugate of $\scal$, and $m_p$ is the mass of the probe. 
We can define the Hamiltonian density from this action in the usual way, giving
\beq
\hat{\cH}=\frac{\Pi^*\Pi}{\rg g^{00}}-\rg g^{ij}\p_i\scal^*\p_j\scal+\rg \,m_p^2\,\scal^*\scal +\ldots
\eeq
where ``$+\ldots$" denotes terms proportional to off-diagonal metric components.

By acting on single particle states (and ignoring the suppressed number changing processes),
we can recast the Hamiltonian in the position representation as a differential operator $H({\bf x},t,-\nabla^2)$ acting on a single particle wave function $\psi({\bf x},t)$. For details of this method, see \cite{Allali:2020shm}.

Specializing to a weakly curved spacetime provided by the DMSMS, with negligible gravitational wave emission, we can use a gauge in which the metric is diagonal. Thus we decompose the metric into a flat background $\eta_{\mu\nu}=\textrm{diag}(+1,-1,-1,-1)$ and a small (diagonal) perturbation $h_{\mu\nu}$ with $|h_{\mu\nu}|\ll 1$ as $g_{\mu\nu}=\eta_{\mu\nu}+h_{\mu\nu}$. Further, 
for static or spherically symmetric sources, we can take $h^{00}=2\Phi(\x,t)$ and $h^{ij}=2\Psi(\x,t) \delta^{ij}$, giving the relatively simple Hamiltonian
\beq
H(\x,t,-\nabla^2)=\sqrt{-\nabla^2+m_p^2}\left(1+\Phi
+\frac{\Psi\,\nabla^2}{\nabla^2-m_p^2}\right)
\label{Hphi}
\eeq
where the first term $\sqrt{-\nabla^2+m_p^2}\equiv H_0$ is the relativistic kinetic energy, and thus we refer to the rest of $H=H_0+V$ as the potential energy  $V(\x,t,-\nabla^2)$.
The Einstein equations give $\nabla^2\Psi=4\pi G_N T_{00}$, and writing $\Phi=\Psi+\delta$, $r(\delta'/r)'= 8\pi G_N(T_{\theta\theta}/r^2-T_{rr})$ for spherically symmetric sources.
The nonrelativistic limit of $H$ gives the familar form $H_{nr}(\x,t,-\nabla^2)=m_p - \nabla^2/2m_p+m_p\Phi$ where the leading $m_p$ is just the constant mass-energy ($m_pc^2$).

{\em Oscillating scalar source}.---
When studying time dependent fields as sources, we can consider coherently oscillating scalar fields as the DMSMS, such as axions which may make up some or all of the missing DM in the universe. As a simple example, the scalar field may locally be characterized by a spherically symmetric spatial profile $\phi_s(r)$ with a single harmonic oscillation frequency $\oma$
\beq \phi(r,t) = \sqrt{2}\,\scprof(r) \cos(\oma t+\varphi)\label{phiosc}
\eeq
For nonrelativistic axion DM, this single harmonic is a realistic approximation with a frequency close to the axion's mass $m_a$. This assumption is also appropriate for condensates of scalars since they have almost perfectly periodic oscillations.
From the Einstein field equations, one can show that the potential based on the choice of eq.~\ref{phiosc} takes the form
\beq 
V(r,t)=\A(r)+\B(r)\cos(2(\oma t+\varphi))\label{VAB}
\eeq

Note that if the axion is nonrelativistic and if the spatial profile of the axion is slowly varying, the (time-averaged) pressure of the source is negligible and $\Phi\approx \Psi$.
In addition, since the frequency $\oma\approx\ma$,
the metric is almost constant in time. For relativistic configurations, the metric can have important time dependence.

{\em Scattering in general}.---The solution of the Schr\"odinger equation $i\partial\psi/\partial t=H\psi$ for the probe particle in a weak gravitational interaction can be studied perturbatively by decomposing into an unscattered part $\psi_u$ and a scattered part $\psi_s$. The unscattered wave function solves the free Hamiltonian Schr\"odinger equation $(i\p_t -H_0)\psi_u=0$, while the scattered part, to first-order in perturbation theory, is a solution to the equation
\beq (i\p_t -H_0)\psi_s(\x,t)=V(\x,t,-\nabla^2)\psi_u(\x,t)\label{linODE}
\eeq
Demanding that the scattered piece $\psi_s$ vanishes in the past, the relevant solution is readily obtained in terms of the retarded Green's function $G_4$ as $\psi_s(\x,t)=\int d^4 x'\, G_4 \,V(\x',t',-\nabla'^2)\psi_u(\x',t')$.

{\em Wave packets}.---We take the incoming wave function to be a wave packet, which is a sum of plane wave modes $\psi^{(q)}(\x,t)=e^{-iE_q t} e^{i\q\cdot\x}$ with amplitudes given by the distribution $\psik(\q)$ peaked at a central value $\k$. The spatial wave function at early times is
\beq 
\psi_u(\x)=\int \!\frac{d^3q}{(2\pi)^3}\frac{1}{\sqrt{2\Eq }}\psik(\q) e^{i\q\cdot\x}e^{-i\q\cdot\bv}
\eeq
where $\bv$ is an impact parameter vector which shifts the center of the wave packet away from the origin of the coordinate system. The factor of $1/\sqrt{2E_q}$ is convenient to make the normalization condition simple $\int d^3q\, |\psik(\q)|^2/(2\pi)^3 =1$.

{\em Scattering from a static source}.---When the potential $V$ is independent of time, and the source of scattering is sufficiently local, one can use the asymptotic form of the retarded Green’s function to solve eq.~\ref{linODE} and write the scattered response of the $q^{\text{th}}$ mode as
\beq\psi_s^{(q)}(\x,t)=e^{- i \Eq  t}f(\q',\q)\frac{e^{iq|\x|}}{|\x|}\label{qthmode}
\eeq
where the scattering amplitude is found to be
\beq
\!\!\!\!\!f(\q',\q)\equiv -\frac{1}{2\pi}\!\int \! d^3 x' e^{i(\q-\q')\cdot\x'}\Big[\Phi(\x')E_q^2+\Psi(\x')q^2\Big]\label{staticamp}
\eeq
Then, the full scattered wave packet is given by
\beq 
\!\!\!\psi_s(\x,t) = \int \frac{d^3q}{(2\pi)^3}e^{-i\Eq  t}f(\q',\q)\frac{e^{iq|\x|}}{|\x|}\frac{ \psik(\q)}{\sqrt{2\Eq }}e^{-i\q\cdot\bv}
\label{psisstat}
\eeq

{\em Scattering from a time-dependent source}.---
We restrict this analysis to coherently oscillating sources with time dependence as in eq.~\ref{VAB}. 
The oscillations of the source can be incorporated into the time dependence of the $q^{\text{th}}$ mode,
allowing a solution similar to eq.~\ref{qthmode} in terms of the appropriately shifted energy variables $E_{\pm1}\equiv E_q\pm 2\omega$
\ba
\psi_s(\x,t)
\amp\,\,=\,\,\amp\sum_{\alpha=0,-1,+1}\int \frac{d^3q}{(2\pi)^3}e^{-i(E_\alpha t+2\alpha\varphi)}\F_\alpha(\q_\alpha',\q)\nonumber\\
\amp\amp\times\frac{e^{iq_\alpha r}}{r}\frac{\psik(\q)}{\sqrt{2\Eq }}e^{-i\q\cdot\bv}\label{psisTD}
\ea
with a set of scattering amplitudes $\F_\alpha$ given by
\ba
\F_0(\q',\q)\amp\,\,\equiv\,\,\amp -\int d^3x' \frac{E_q}{2\pi}e^{i(\q-\q')\cdot\x'}\A(r',q^2)\\
\F_{\pm1}(\q_{\pm1}',\q)\amp\,\,\equiv\,\,\amp -\int d^3x' \frac{E_{\pm1}}{4\pi}e^{i(\q-\q_{\pm1}')\cdot\x'}\B(r',q^2)\,\,\,\,\,\,\,\label{Fdef}
\ea
and shifted momenta $q_{\pm1}\equiv (E^2_{\pm1}-m_p^2)^{1/2}$.

{\em Decoherence}.---Interactions ensure that any macroscopic quantum superposition and its environment become inevitably entangled. Thus, the environment will evolve into a corresponding superposition. Since the degrees of freedom of the composite system become numerous as the interactions continue, an observer cannot track the full detail of the system. Taking on a coarse-grained point of view, the observer will ignore the environment's degrees of freedom and trace them out of the density matrix. This effectively spoils the quantum coherence of the remaining sub-system.

We now apply the scattering formalism to a probe particle which interacts with a DMSMS that begins in a superposition of otherwise classical states $\ket{\DM_1}$ and $\ket{\DM_2}$. This generates a potential for the probe which is in a superposition of potentials $V_1$ and $V_2$, and the probe evolves into a superposition of
$\ket{\psi_1}$ and $\ket{\psi_2}$. This evolution is presented pictorially in fig.~\ref{scheme}. 

{\em Decoherence rate}.---The decoherence rate depends on the inner product of the sub-states $\ket{\psi_1}$ and $\ket{\psi_2}$ which we can parameterize by its deviation from unity $|\braket{\psi_1}{\psi_2}|\equiv 1-\Delta$. The inner product consistent with the relativistic normalization conditions is given by $\braket{\psi}{\phi}\equiv\int d^3 x \left(-i \phi(\x,t)\p_t \psi^*(\x,t)+i\psi^*(\x,t)\p_t\phi(\x,t)\right)$. To lowest order in scattering, the leading contributions to $\Delta$ come from the first-order scattered wave function defined previously. One can show that, at this order, $\Delta$ is approximately \cite{Allali:2020ttz}
\beq 
\!\Delta=\frac{1}{2}\left( \braket{\psi_{s,1}}{\psi_{s,1}}+\braket{\psi_{s,2}}{\psi_{s,2}}-2\,\Re\!\left[\braket{\psi_{s,1}}{\psi_{s,2}}\right]\right)\label{Delta}\eeq
Then, the decoherence rate is found by summing
the $\Delta$'s from the many probe particles
\beq
\Gamd=n\,v\int \! d^2b\,\Delta_b
\label{DecRate}\eeq
where $n$ is the number density of probe particles, $v$ is their typical speed, and $\Delta_b$ is $\Delta$ evaluated at impact parameter ${\bv}$.
Thus, the following integral over impact parameter is crucial
\beq
S_{ij}\equiv \int d^2 b \braket{\psi_{s,i}}{\psi_{s,j}}\label{SijDef}
\eeq
which is used to give the decoherence rate as
\beq 
\Gamd= \frac{1}{2} n v (S_{11}+S_{22}-2\Re[S_{12}])
\label{Gamdsimp}
\eeq

For a more general DM superposition $\sum_i a_i \ket{\DM_i}$, the overlaps $\braket{\psi_i}{\psi_j}$ still control the decoherence rate. If these overlaps are of a similar order, the inferred rate will be similar to that of a two-component superposition. Thus, we believe this simplification to be justified.

{\em Decoherence from a static source}.---We can obtain an expression for the decoherence rate by first inserting eq.~\ref{psisstat} into eq.~\ref{SijDef}, where $\psi_i$ and $\psi_j$ are written in terms of momenta $\q$ and $\qtv$, respectively.
We include the addition shifts $\bv\to\bv-\Lv_i$ to indicate that the centers of the sources $V_i$ may differ.
The integral can be readily simplified when the distributions $\psik$ are narrowly peaked around $\k$ (for details see \cite{Allali:2020shm}). Averaging over the direction of $\k$ replaces the $\psik$ with a function $P_k(q)$ which depends only on the magnitudes $k$ and $q$, giving simply
\beq S_{ij}=\int \frac{d^3 q}{(2\pi)^3}\sigmat_{ij}(q)P_k(q) \eeq
where we have defined the generalized scattering cross section $\sigmat_{ij}$ to be
\beq
\!\!\!\!\sigmat_{ij}(q)\equiv\int \!d^2\Omega\,f_i^{*}(\q',\q)f_j(\q',\q)\,j_0(2q L_{ij}\sin(\theta/2))\label{sigmatilde}
\eeq
where $j_0(z)\equiv \sin(z)/z$ is the sinc-function and $L_{ij}\equiv|{\bf L}_i-\Lv_j|$. Note that $\sigma_1\equiv\sigmat_{11}$ and $\sigma_2\equiv\sigmat_{22}$ are the usual scattering cross sections (with the appropriate general relativistic amplitudes of eq.~(\ref{staticamp})) since $j_0(0)=1$.

If the $\psik$ are narrow enough, we may approximate $\q\approx\k$ and take $S_{ij}\approx\sigmat_{ij}$, obtaining the result
\beq 
\Gamd= \frac{1}{2} n\, v (\sigma_{1}+\sigma_{2}-2\Re[\sigmat_{12}])|_{q=k}\label{GamdecStatic}
\eeq
Note that in the special case when $\sigmat_{12}=0$ and $\sigma\equiv\sigma_1=\sigma_2$, we recover the familiar form $\Gamd = n\, v \,\sigma$ relating to other results in the literature on decoherence. However, our result in eq.~(\ref{GamdecStatic}) generalizes this to include the non-trivial cross term $\sigmat_{12}$ and full relativistic corrections.

{\em Decoherence from a time-varying source}.---An analogous procedure may be applied to compute first $S_{ij}$ and then the decoherence rate for time-varying sources. Each $S_{ij}$ involves a nine-term sum over the $\alpha$'s of each sate $\psi_i$ and $\psi_j$; for a detailed analysis, see \cite{Allali:2020shm}. In particular, considering narrow $\psik$ will be of interest once more, allowing us to simplify the decoherence rate in terms of another generalized cross section $\sig$, defined in eq.~\ref{Sigmadef}.

{\em Phase difference}.---We examine now the case where the distinction between the $V_i$
is only in the phase of oscillation $\varphi_i$. Thus we set $\Lv_1=\Lv_2=0$, and the overlap integral $S_{ij}$ can be greatly simplified
\beq S_{ij}=\sum_{\alpha,\beta}\sig_{\alpha\beta}\,e^{2i(\alpha\varphi_i-\beta\varphi_j)}\label{Ssum}\eeq
where the sum runs over $\alpha,\beta=-1,0,+1$, we have made the approximation that the distributions $\psik$ are narrowly peaked to integrate over momentum, and we have defined the new generalized cross sections as
\beq
\sig_{\alpha\beta}(\k)\equiv \int d^2\Omega\,\F^*_\alpha(\k_\alpha',\k)\F_\beta(\k'_{\alpha},\k_{\alpha-\beta})\label{Sigmadef}
\eeq
Using the fact that $\Re[\sig_{\alpha\beta}]=\Re[\sig_{\beta\alpha}]$ (see \cite{Allali:2020shm}), we can see that the terms $\sig_{0,0}$, $\sig_{0,\beta}$, and $\sig_{\alpha,0}$ drop out of the expression for the decoherence rate in eq.~\ref{Gamdsimp}, giving
\beq\Gamd= \frac{1}{2} n v \Big( S_\varphi(\sig_{+1,+1}+\sig_{-1,-1})-A_\varphi \Re[\sig_{-1,+1}]\Big)\label{Gamdsphi}
\eeq
where $S_\varphi=2(1-\cos(2(\varphi_1-\varphi_2)))$ and $A_\varphi=\cos(4\varphi_1)+\cos(4\varphi_2)-2\cos(2(\varphi_1+\varphi_2))$ encode the phase dependence. This is an interesting new quantum gravitational result.

{\em Quantitative results}.---
For situations where the superposition states differ in their spatial profiles, the nonrelativistic limit of our formalism usually suffices, compatible with \cite{Allali:2020ttz}.
If we consider a DMSMS which is just a random Gaussian overdensity surrounded by an underdensity in the background density of DM in the galaxy (with vanishing monopole), we can set the size of the DMSMS by the de Broglie wavelength of typical DM in the Milky Way and obtain the following decoherence rate
$\Gamd = CG_N^2m_p\rhoDM^2\rho_p/(m_a^8 v_a^8 v_p)$, 
where $\rhoDM$ and $\rho_p$ are the densities of the DM and the probe, $v_a$ and $v_p$ are the typical speeds of the DM and probe particles, and $C$ is a constant numerical factor (for details, see \cite{Allali:2020ttz}). 
Estimating this rate using the local density of DM in the Milky Way $\rho_{loc,mw} \approx 0.4 \,\mbox{GeV}/\mbox{cm}^3$ \cite{Read:2014qva}, taking the probe to be a proton with density $\rho_p\sim 0.2\, \rho_{loc,mw}$, and estimating the speed of axions and protons as the virial speed in the Milky Way $v_p\approx v_a\approx 220 \,\mbox{km/s}$, we find that $\Gamd \approx 10^{-20}\,\mbox{sec}^{-1} (10^{-6}\,\mbox{eV}/\ma)^8$. This means that light axions decohere rapidly. We can also examine decoherence of such a DMSMS if it passes through the earth's atmosphere; using the density of probes to be $\rho_p\sim 1\, \mbox{kg/m}^3$, we find $\Gamd \approx 10^{3}\,\mbox{sec}^{-1} (10^{-6}\mbox{eV}/\ma)^8$, which can be significant.

For the remainder of this work, we will focus on the 
above
truly relativistic phenomenon when the superposition states only differ by the phase of the axion. In this case, there is no divergence in the forward direction in the scattering amplitudes since the only cross sections left in eq.~\ref{Gamdsphi} do not diverge even for a monopole (due to the fact that the transfer momentum after scattering cannot 
vanish when $\alpha,\beta\neq0$). 

We will choose the spatial profile of the DMSMS to be a Gaussian. 
A Gaussian perturbation in the DM density corresponds to a Gaussian momentum distribution, fitting the expectation that virialized DM in the galaxy would obey a Maxwell velocity distribution. 
We take the spatial profile to be  
$\phi_s(r)\sim\sqrt{\kappa M\mu^3/\ma^2}\,e^{-r^2\mu^2/2}$, 
where $M$ is the mass-scale associated with the DMSMS, $\mu$ is the inverse length scale (roughly denoting its size), and $\kappa$ is an $\cO(1)$ numerical factor. The scattering amplitudes in eq.~\ref{Fdef} are proportional to Fourier transforms of the potential with respect to the transfer momentum $\pv_{tr}\equiv\q-\q_\alpha$. Therefore, the cross sections and the decoherence rate will be proportional to the Gaussian $e^{-\pv_{tr}^2/4\mu^2}$. The transfer momentum is a minimum in the forward direction, and for $\oma\ll E_q$ (which is expected for realistic probes and an axion DMSMS), the argument of the exponential is $\pv_{tr}^2/(4\mu^2)\approx\alpha^2\oma^2 E_q^2/(\mu^2 q^2)$ (see \cite{Allali:2020shm}). For the relevant terms $\alpha\neq0$, this cannot vanish; thus
\beq
\Gamd\propto \exp\left[-{\omega^2E_k^2\over\mu^2k_p^2c^4}\right]
\eeq
(reinstating $c$). So if the transfer momentum is appreciable, the decoherence rate 
is exponentially suppressed.

{\em Slowly moving dark matter}.---For slow axion DM, one expects $\oma\approx\ma\,c^2$. We can then set the scale $\mu$ by the de Broglie wavelength of the axion such that $\mu\sim p_a=m_a v_a$.
Taking the probe to be a proton in the galaxy, we set the speeds of both particles to be the virial speed in our galaxy $\sim 10^{-3} c$, obtaining
\beq 
{\omega^2E_k^2\over\mu^2k_p^2c^4}\approx\frac{\oma^2 (m_p^2 + k_p^2/c^2)}{m_a^2 v_a^2 k_p^2}\approx\frac{c^4}{v_a^2 v_p^2}\sim 10^{12}
\eeq
Similarly, if the probe is relativistic, such as a photon, we find
$\omega^2E_k^2/(\mu^2k_p^2c^4)\approx c^2/v_a^2\sim 10^6$.
Thus, for nonrelativistic objects in the galaxy, the decoherence rate will be exponentially suppressed and a superposition of the phase of the axion is robust against decoherence.

{\em Relativistic dark matter}.---If instead we consider a component of the DM that is relativistic, such as near a black hole or for dense boson stars (see below), we can have situations where $\oma\sim\mu\,c$ and thus the exponential may not necessarily suppress the decoherence rate. 

First, let us examine the dependence of the decoherence rate on the physical scales (see \cite{Allali:2020shm} for detailed analysis)
\beq 
\Gamd
\approx K\frac{G_N^2 E\, \rho_p\, \rhoDM^2}{\ma^8}\,\,\,\,\,\,[\mbox{phase; (semi)-relativistic}]
\label{Gammasimple}\eeq
(here $c=1$), where $E$ is the energy of the probe and $K$ is a numerical factor depending on the ratios $k_p/m_p$, $\oma/\ma$, and $\mu/\ma$ (each is $\cO(1)$ or greater for DM and probes which are relativistic). 

To obtain an idea of the decoherence rate, we shall use the local average density in the Milky Way $\rho_{loc,mw}$ for the density of the DMSMS. If the probes are semi-relativistic protons in the galaxy, we can take their energy to be $\sim 2\,\mbox{GeV}$. Such protons may come from cosmic rays, so we may use estimates of cosmic ray proton density from the literature \cite{Persic:2017qxo}, roughly $\rho_{cr}\sim 10^{-9}\,\mbox{GeV}/\mbox{cm}^3$. Finally, taking a representative $K$ to be $10^{-4}$ for semi-relativistic DM and protons, we find a reference decoherence rate near the Hubble rate today ($H_0\approx2.2\times10^{-18}\,\mbox{s}^{-1}$) to be 
$\Gamd\sim 10^{-21}\,\mbox{sec}^{-1}(K/10^{-4})(10^{-12}\,\mbox{eV}/\ma)^8$.
Similarly, we can consider photons from the cosmic-microwave background (CMB) as probes of the DMSMS, which have number densities of approximately $n\sim 400/\mbox{cm}^3$ and typical energies of $\sim6\times10^{-4}\,\mbox{eV}$. Taking a representative value of $K$ for the semi-relativistic DM with a photon probe to be $10^{-2}$, we find
the decoherence rate  $\Gamd\sim 10^{-16}\,\mbox{sec}^{-1}(K/10^{-2})(10^{-14}\,\mbox{eV}/\ma)^8$.

{\em Boson stars}.---Axions are predicted to form gravitationally bound Bose-Einstein condensates known as boson stars. Previous work has established the dynamics of boson stars \cite{Tkachev:1986tr,Gleiser:1988rq,Seidel:1990jh,Tkachev:1991ka,Jetzer:1991jr,Kolb:1993zz,Liddle:1993ha,Sharma:2008sc,Chavanis:2011zi,Chavanis:2011zm,Liebling:2012fv,Visinelli:2017ooc,Schiappacasse:2017ham,Hertzberg:2018lmt,Hertzberg:2018zte,Levkov:2018kau}, and specifically the unique phase-dependence of the outcomes of boson star mergers \cite{Hertzberg:2020dbk}. Considering a typical dilute boson star made of nonrelativisitc axions, if it is in a superposition of phases, we can predict that the coherence of the superposition will be long-lived. Further, if such a boson star engages in a merger event, the phase dependence would lead to non-trivial evolution of the merged object. 

In contrast, the densest boson stars \cite{Helfer:2016ljl} are 
made of (semi)-relativistic axions, and therefore they may 
exhibit appreciable decoherence of their phases.
We find the decoherence rate to be $\Gamd \sim 10^6 \,\mbox{sec}^{-1}(K/10^{-4})(1\,\mbox{eV}/\ma)^4$, which is quite rapid. 

{\em Black holes}.---
We also remark that rapid decoherence for DM would occur near the horizon of a black hole as the DM becomes highly relativistic. It is interesting to note that from this point of view, the most classical states are entering the black hole; this may have ramifications for the information paradox.

{\em Consequences for earth based experiments}.---Earth-based experiments, for example haloscopes \cite{Sikivie:1983ip} including ADMX \cite{Du:2018uak}, search directly for axion DM that passes through the atmosphere or the earth. We previously showed that the decoherence rate is increased by frequent interactions when the DM passes through the atmosphere \cite{Allali:2020ttz}. However, the phase of the axion cannot be decohered in this way for a nonrelativistic DMSMS. Since the DM near the earth should be nonrelativistic, 
our results show that 
a superposition of phases 
will survive. 

This suggests that earth-based experiments
should consider a 
quantum superposition
of axion waves of different phases interacting with the detector
$|\mbox{axion} \rangle \sim \sum_i c_i |\cos(\omega t - {\bf k}_a\cdot{\bf x}+\varphi_i)\rangle$.
Experiments like ADMX involve a resonant cavity, and the phase of the resonant electromagnetic waves will be impacted by the phase of the axion. This raises questions about what impact
this
may have for the signal in the resonant cavity. The cavity would evolve into a superposition of different resonant electromagnetic signals, though it may be difficult to predict the experimental signature of this phenomenon, since the subsequent decoherence from other interactions will likely be very rapid. 
This fundamentally new phenomenon found here can alter the response of detectors and deserves consideration in detection strategies.

We assumed in this work that a DM superposition may naturally exist. One may also attempt to determine whether some DM is in a superposition by probing it with a particle which is entangled with a reference system; the probe particle, upon remeasurement, may divulge information about the state of the DM. An individual particle generally only gains a small amount of entanglement with DM, and it is the net effect of a large number of particles that can lead to decoherence. Nevertheless, this may be an interesting way to learn about the character of the DM state.

\begin{acknowledgments}
M.~P.~H. is supported in part by National Science Foundation Grant No. PHY-2013953.

$^*$itamar.allali@tufts.edu, $^\dagger$mark.hertzberg@tufts.edu
\end{acknowledgments}

\end{document}